\documentclass[%
 reprint,
 amsmath,amssymb,
 aps,
]{revtex4-1}

\usepackage{graphicx}% Include figure files
\usepackage{dcolumn}% Align table columns on decimal point
\usepackage{bm}% bold math
\begin{document}

\preprint{APS/123-QED}

\title{A parameter free similarity index based on clustering ability \\for link prediction in complex networks}% Force line breaks with \\

\author{Zhihao Wu }
\email{zhwu@bjtu.edu.cn}
 
\author{Youfang Lin}

\author{Yao Zhao}
\affiliation{%
 Beijing Key Lab of Traffic Data Analysis and Mining,\\
 School of Computer and Information Technology,\\
Beijing Jiaotong Univerisy
}%

%\date{\today}% It is always \today, today,
             %  but any date may be explicitly specified

\begin{abstract}
Link prediction in complex network based on solely topological information is a challenging problem. In this paper, we propose a novel similarity index, which is efficient and parameter free, based on clustering ability. Here clustering ability is defined as average clustering coefficient of nodes with the same degree. The motivation of our idea is that common-neighbors are able to contribute to the likelihood of forming a link because they own some ability of clustering their neighbors together, and then clustering ability defined here is a measure for this capacity. Experimental numerical simulations on both real-world networks and modeled networks demonstrated the high accuracy and high efficiency of the new similarity index compared with three well-known common-neighbor based similarity indices: CN, AA and RA. \\
\\
\begin{description}
\item[PACS numbers]
89.75.Hc, 89.20.Hh
\end{description}
\end{abstract}

\pacs{Valid PACS appear here}% PACS, the Physics and Astronomy
                             % Classification Scheme.
%\keywords{Suggested keywords}%Use showkeys class option if keyword
                              %display desired
\maketitle

%\tableofcontents

\section{\label{sec:level1}INTRODUCTION}

Many complex systems can be modeled using complex networks, such as social, biological and information systems, and the study of complex networks has attracted increasing attention and becomes a popular tool in many different branches of science \cite{albert2002statistical,boccaletti2006complex,costa2007characterization,bianconi2009assessing,shen2014collective}. Link prediction in complex networks aims at estimating the likelihood of the existence of a link between two nodes, and it has many applications in different fields. For example, predicting whether two users know each other can be used to recommend new friends in social networking sites, and in the field of biology, accurate prediction of protein-protein interaction has great value to sharply reduce the experimental costs, especially when our knowledge is very limited. For example, 80\% of the molecular interactions in cells of Yeast \cite{yu2008high} and 99.7\% of human \cite{stumpf2008estimating} are still unobserved. Large amount of missing links make the observed networks sparser, which of course leads to more difficulty to predict. 

The problem of link prediction can be defined in different backgrounds considering different information. In this paper, we focus on link prediction relying on solely topological information. There are three main kinds of methods: local, global and quasi-local methods \cite{liben2007link,lu2011link}. Local methods \cite{lorrain1971structural,adamic2003friends,zhou2009predicting,cannistraci2013link} are always very efficient, while global ones \cite{katz1953new,jeh2002simrank,gobel1974random,fouss2007random} are more accurate. Quasi-local methods, such as Ref.~\cite{lu2009similarity}, can be designed by adding some constraints on global ones, but always bring some parameters. 

CN \cite{lorrain1971structural} is the most well-known local similarity index, which regards pair of nodes with more common neighbors is more likely connected by a link. The drawback of CN is that all common neighbors are treated as the same, so that many pairs of nodes are given the same likelihood. AA \cite{adamic2003friends} and RA \cite{zhou2009predicting} indices solve this problem by distinguishing common-neighbors by node-degree in different way. Some prediction results have shown AA and RA can predict missing links more accurately than CN for better resolution. But in some cases, degree is still limited. For example, the max degree of some large sparse networks is not very big, so degree is not able to provide sufficient resolution in these cases. In this paper, we will focus on this problem and give a novel similarity index with better discriminative resolution.

Besides, there are also some sophisticated models to solve the problem of link prediction. Clauset et al. proposed an algorithm based on the hierarchical network structure, which gives good predictions for the networks with hierarchical structures \cite{clauset2008hierarchical,redner2008networks}. Guimera et al. solved this problem using stochastic block model \cite{guimera2009missing}. Recently, Linyuan L\"{u} et al. propose a concept of structural consistency, which can reflect the inherent link predictability of a network, and they also propose a structural perturbation method for link prediction, which is more accurate and robust than the state-of-the-art methods \cite{lu2015toward}. Not only these methods can give good prediction results in some networks, another significance of these methods is to give insights into the mechanism of link formation, network evolution, and even the link predictability \cite{lu2015toward}. However, there should be more effort to make these methods efficient enough.

In this paper, we will propose a simple, efficient and parameter free similarity index based on clustering ability, which is defined by average clustering coefficient of nodes with the same degree. The study of Liben-Nowell and Kleinberg showed that CN and AA perform better than other seven well-known local similarity indices. Some later literature showed that RA perform even better than CN and AA. So we compare our method with the three state-of-the-art common-neighbor based indices in experiments. Experimental results on 12 real-world networks drawn from 6 different fields show that our method outperforms other compared well-known common-neighbor based methods. Especially, we find the new similarity index can always give impressive promotion in sparse networks with low average clustering coefficient. Further, we verify this point employing a tunable network model.

\section{\label{sec:level1}METHODS}
Considering an unweighted undirected simple network $G(V,E)$, where $V$ is the set of nodes and $E$ is the set of links. For each pair of nodes, $x,y\in V$, we assign a score $s_{xy}$. Since $G$ is undirected, the score is symmetry. All the nonexistent links are sorted in decreasing order according to their scores, and the links in the top are most likely to exist. The common-used framework always sets the similarity to the score, so the higher score means the higher similarity, and vice versa.

\subsection{Compared similarity indices}
In this paper, we will compare our method with three well-known similarity indices: CN, AA and RA. Their definitions and relevant motivations are introduced as follows:

(1)	CN (common neighbors) For a pair of nodes, CN counts the number of common neighbors. In common sense, more common neighbors indicates larger probability to form/exist a link between two nodes. The definition of CN is given in equation (\ref{eq1}), in which $\Gamma(x)$ denotes the set of neighbors of node $x$.

\begin{equation}
s_{xy}^{CN} = |\Gamma(x)\cap\Gamma(y)|
\label{eq1}
\end{equation}

(2)	AA (Adamic-Adar). AA refines the simple counting of common neighbors by assigning the less-connected nodes more weight, and is defined as equation (\ref{eq2}).
\begin{equation}
s_{xy}^{AA} = \sum_{z\in|\Gamma(x)\cap\Gamma(y)|}\frac{1}{log(k_z)}
\label{eq2}
\end{equation}

(3)	RA (Resource allocation). RA is also an index based on common neighbors, and the motivation comes from resource allocation dynamics on complex systems. For a pair of unconnected nodes, $x$ and $y$, the node $x$ can send some resource to $y$, with their common neighbors playing the role of transmitters. In the simplest case, assume that each transmitter has a unit of resource, and will equally distribute the resource to all its neighbors. The similarity can be defined as given by equation (\ref{eq3}), which measures the amount of resource $y$ received from $x$. Comparing with AA (Adamic-Adar), which only simply replaces $k_z$ by $log(k_z)$, the little difference only makes the results different significantly when the degrees of common neighbors are comparatively high. 
\begin{equation}
s_{xy}^{RA} = \sum_{z\in|\Gamma(x)\cap\Gamma(y)|}\frac{1}{k_z}
\label{eq3}
\end{equation}

In respect to CN, most other CN-based similarity indices are design by weakening high-degree nodes or bring some other link or structural information into the definition of measures. In one word, all these motivations can be summed up in offering more discriminative resolution. 

\subsection{The new similarity index}
The new index is called CA (clustering ability) and its definition is given in equation (\ref{eq4}).
\begin{equation}
s_{xy}^{CA} = \sum_{z\in|\Gamma(x)\cap\Gamma(y)|}C(k_z)
\label{eq4}
\end{equation}

where $C(k_z)$ is the average clustering coefficient of nodes with degree equal to $k_z$. The clustering coefficient of a node is defined in equation (\ref{eq5}).
\begin{equation}
C_i = \frac{2t_i}{k_i(k_i-1)}
\label{eq5}
\end{equation}
where $t_i$ is the number of triangles passing through node $i$ and $k_i$ is the degree of node $i$.

The motivation of the CA index comes from the assumption that common-neighbors can contribute likelihood to a pair of nodes, because they have some ability to cluster other nodes together. Please note clustering ability has some difference with the current clustering situation of a node. For example, we cannot say a node with only two neighbors that are not connected has no ability to cluster its neighbor together. On the contrary, if the link between its neighbors is unobserved or missing, we will totally make a wrong judgement. Thus, we use the average clustering coefficient of nodes with the same degree to estimate the clustering ability of a node, and we think it is a more robust way.

One may ask whether there is a big difference between $C(k)$, $1/k$ and $1/log(k)$. The answer is definitely yes. We plot $C(k)$ versus $k$ for all tested networks in Figure~\ref{fig1} and Figure~\ref{fig2}. It is easy to find in most cases their distributions on $k$ are very different. There are two main aspects: 1) when degrees are relatively small, $C(k)$ not only does not decrease as fast as $1/k$ or $1/logk$, but even increases a bit for some networks, such as PPI1 and PPI2 networks. In most cases, $C(k)$ distributes a little flat or decreases very slow when $k$ is small; 2) when degrees are large, the distribution of $C(k)$ can be very broad, instead of a straight line (RA) or a curve (AA). That means nodes with similar degrees can provide very different contributions in forming a link.

\subsection{Estimation}
To estimate the predicted results comprehensively, here we employ two estimators: AUC and precision, which are commonly used in other related literatures \cite{lu2015toward,lu2009similarity}. The basic preparation for calculation of the two methods is the same. To test the precision of a prediction algorithm, the observed links $E$ is randomly divided into two parts: the training set $E_t$ is treated as known information, while the probe set $E_p$ is used for testing and no information in the probe set is allowed to be used for prediction. Obviously, $E = E_t \cup E_p$ and $E_t \cap E_p = null$. In this paper, we consider 10\% of links as test links.

AUC is a standard metric, the area under the receiver operating characteristic (ROC) curve, to quantify the accuracy of the prediction algorithms. In this situation, it can be interpreted as the probability that a randomly chosen missing link (belongs to $E_p$) is given a higher score than a randomly chosen nonexistent link (which belongs to $U-E$, where $U$ denotes the set of all node-pairs). In practice, the calculation of AUC is given as defined by equation (\ref{eq6}), where $n$ is the times of independent comparisons, $n'$ denotes the times the missing links having a higher score, and $n''$ counts situations they have the same score. A higher value of AUC indicates better results.

\begin{equation}
AUC = \frac{n'+0.5n''}{n}
\label{eq6}
\end{equation}

Given the ranking of the non-observed links, the precision is defined as the ratio of relevant items selected to the number of items selected. That means if we take the top-L links as the predicted ones, among which $L_r$ links are right, then the precision can be defined as equation (\ref{eq7}). Higher precision indicates higher prediction accuracy.

\begin{equation}
precision = \frac{L_r}{L}
\label{eq7}
\end{equation}

\section{EMPIRICAL ANALYSIS}
\subsection{Tests on real-world networks}
In this paper, we will compare CA index with three other well-known common-neighbor based similarity indices on 12 real-world networks drawn from various fields. PPI1 \cite{ben2005kernel} and PPI2 \cite{ben2005kernel,chen2006increasing} are two protein-protein interaction networks. Food \cite{cohen2009food} and Grassland \cite{dawah1995structure} are two food web networks. Dolphins \cite{lusseau2003bottlenose} and Jazz \cite{gleiser2003community} are dolphins and musician social networks. MacNeu \cite{kotter2004online} and MouseNeu \cite{bock2011network} are two neural networks. PB \cite{adamic2005political} and Email \cite{guimera2003self} are two social networks from electronic information systems. Grid \cite{watts1998collective} and INT \cite{spring2002measuring} are two artificial infrastructure networks. The basic topological features of these networks are given in Table~\ref{t1}.

\begin{table}[h!]
\caption{The basic topological features of the 12 real-world networks. N and M are the total number of nodes and links, respectively. \textless k\textgreater is the average degree of the network. \textless d\textgreater is the average shortest distance between node pairs. \textless C\textgreater  is the average clustering coefficient.}
\begin{ruledtabular}
\begin{tabular}{llllll}
\multicolumn{1}{l}{\textbf{Nets}}&\multicolumn{1}{l}{\textbf{N}} & \multicolumn{1}{l}{\textbf{M}} 
&\multicolumn{1}{l}{\textbf{\textless k\textgreater}} & \multicolumn{1}{l}{\textbf{\textless d\textgreater}} & \multicolumn{1}{l}{\textbf{\textless C\textgreater}}                                                                                                     \\ \hline
PPI1      & 4036  & 10411  & 5.159  & 4.412  & 0.0682     \\ 
PPI2      & 4385  & 12234  & 5.58   & 4.424 & 0.0911     \\ 
Food      & 51    & 233    & 9.137  & 2.063  & 0.0813     \\ 
Grassland & 75    & 113    & 3.013  & 3.875  & 0.3377     \\ 
Dolphins  & 62    & 159    & 5.129  & 3.357  & 0.259      \\ 
Jazz      & 198   & 2742   & 27.7   & 2.235  & 0.6175     \\ 
MacNeu    & 94    & 1515   & 32.23  & 1.771 & 0.7736     \\ 
MouseNeu  & 18    & 37     & 4.111  & 1.967 & 0.2163     \\ 
PB        & 1222  & 16714  & 27.36  & 2.738 & 0.3203     \\ 
Email     & 1133  & 5451   & 9.622 & 3.606  & 0.2202     \\ 
Grid      & 4941  & 6594   & 2.669  & 15.87  & 0.0801     \\ 
INT       & 5022  & 6258   & 2.492  & 5.99   & 0.0116     \\ 

\end{tabular}
\end{ruledtabular}
\label{t1}
\end{table}

\begin{figure*}
\includegraphics{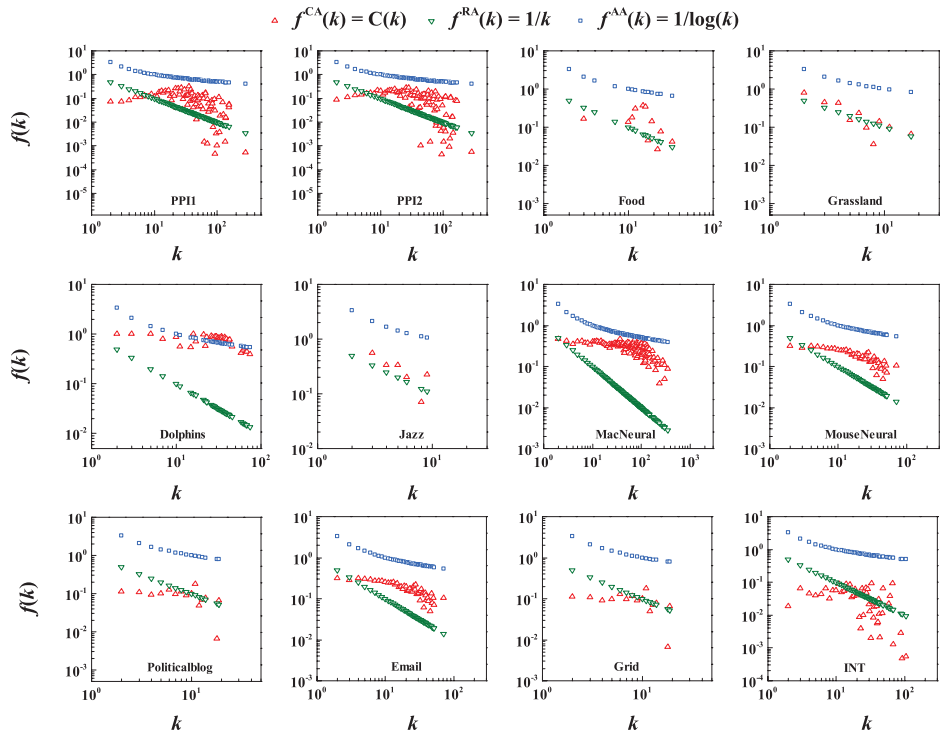}% Here is how to import EPS art
\caption{\label{fig1} Common-neighbor's contribution versus degree for CA, RA and AA in 12 real-world networks.C(k) indicates the average clustering coefficient of nodes with degree k.}
\end{figure*}

In Figure~\ref{fig1}, we plot the corresponding measure of common-neighbor's contribution for CA, RA and AA in the twelve real-world networks. It shows that there are big differences among the three indices, whether for small-degree nodes or large-degree nodes. For nodes with small degree, $C(k)$, which indicates average clustering coefficient of nodes with degree $k$, always decreases very slowly with the increase of degree, and for PPI networks the trend is even in the opposite direction. For nodes with large degree, $C(k)$ always distributes more broad, rather than as a line or a curve.

First, we show the experimental results of the four efficient methods evaluated by AUC in Table~\ref{t2}, with those entries corresponding to the highest accuracies being emphasized by black. All experimental results in this paper are average of 100 runs. RA performs the best among all methods. CA performs the second-best and is slightly worse than RA. On five networks, AA, RA and CA get the same results. On the rest of networks, RA gets four best results and CA gets another three best results. 

\begin{table}[h!]
\caption{Link prediction accuracy measured by AUC on 12 real-world networks.}
\begin{ruledtabular}
\begin{tabular}{lllll}
\multicolumn{1}{l}{\textbf{AUC}}&\multicolumn{1}{l}{\textbf{CN}} & \multicolumn{1}{l}{\textbf{AA}} 
&\multicolumn{1}{l}{\textbf{RA}} & \multicolumn{1}{l}{\textbf{CA}}                                                \\ \hline
PPI1       &0.714   &\bf{0.715}   & \bf{0.715}   &\bf{0.715}   \\ 
PPI2       &\bf{0.738}   &\bf{0.738}   &\bf{0.738}   &\bf{0.738}   \\  
Food       &0.397   &0.409   &0.419   &\bf{0.445}   \\ 
Grassland  &0.782   &0.796   &\bf{0.797}   &0.790   \\  
Dolphins   &0.796   &0.799   &0.797   &\bf{0.800}   \\  
Jazz       &0.956   &0.963   &\bf{0.972}   &0.962   \\  
MacNeu     &0.944   &0.945   &\bf{0.949}   &0.946   \\  
MouseNeu   &0.467   &0.475   &0.476   &\bf{0.483}   \\  
PB         &0.924   &0.927   &\bf{0.929}   &0.928   \\  
Email      &0.856   &\bf{0.858}   &\bf{0.858}   &\bf{0.858}   \\  
Grid       &\bf{0.625}   &\bf{0.625}   &\bf{0.625}   &\bf{0.625}   \\  
INT        &\bf{0.653}   &\bf{0.653}   &\bf{0.653}   &\bf{0.653}   \\  
\end{tabular}
\end{ruledtabular}
\label{t2}
\end{table}

Table~\ref{t3} shows the precision results of the compared indices on the 12 networks. Clearly, CA performs better than all other three indices with a distinct advantage. Among all tested networks, CA can predict more precisely than AA and RA. Only on PB network, CN gets a better result than CA. At the same time, big improvements are attained in many networks, such as PPI1, PPI2, Food, Grid and INT networks. One may note that all the above five networks have two common statistical features: low average clustering coefficient and relatively low average degree. If we look back at these networks in Figure 1, we can find an interesting phenomenon: these networks are exactly the ones in which the distributions of $C(k)$ are very different from $1/k$ and $1/log(k)$. We figure that this goes to show the capacity of CA in better estimating the contribution of common-neighbors. The large improvements may have relations with both low average clustering coefficient and low average degree, although it is hard to give a definite theoretical analysis. We will give some more evidence on this issue employing a network model in the following section.

\begin{table}[h!]
\caption{Link prediction accuracy measured by precision on 12 real-world networks.}
\begin{ruledtabular}
\begin{tabular}{lllll}
\multicolumn{1}{l}{\textbf{Prec}}&\multicolumn{1}{l}{\textbf{CN}} & \multicolumn{1}{l}{\textbf{AA}} 
&\multicolumn{1}{l}{\textbf{RA}} & \multicolumn{1}{l}{\textbf{CA}}                                                \\ \hline
PPI1      &0.184 &0.145 &0.080 &\bf{0.202} \\ 
PPI2      &0.236 &0.196 &0.109 &\bf{0.240} \\  
Food      &0.007&0.006&0.008&\bf{0.023}    \\  
Grassland &0.064&\bf{0.126}&\bf{0.126}&\bf{0.126}\\  
Dolphins  &0.128&0.114&0.095&\bf{0.130}\\ 
Jazz      &0.821&0.838&0.824&\bf{0.850}\\  
MacNeu    &0.574&0.578&0.554&\bf{0.604}\\  
MouseNeu  &0.046&0.047&0.047&\bf{0.052}\\  
PB        &\bf{0.418}&0.380&0.252&0.400\\  
Email     &0.293&0.320&0.256&\bf{0.328}\\  
Grid      &0.120 &0.098 &0.080 &\bf{0.131} \\  
INT       &0.105 &0.104 &0.083 &\bf{0.110} \\  
\end{tabular}
\end{ruledtabular}
\label{t3}
\end{table}

\subsection{Tests with PS model}
To demonstrate the relationship between the advantages of CA index and network features, we test the above four similarity indices on artificial networks generated by Popularity versus Similarity (PS) model \cite{papadopoulos2012popularity}. PS model considers the factor of popularity and similarity at the same time in the growing procedure of a network, so that it can generate networks with more similar features with those of real-world networks than some other well-known models, such as WS model \cite{watts1998collective}, BA model \cite{barabasi1999emergence} and etc. 

Two parameters of PS model are tuned in our analysis: one is the temperature parameter $T$, which can be used to tune the average clustering coefficient of the generated network, and the other one is $m$, which is a parameter controlling the average node degree $\textless k\textgreater = 2m$. The parameter $T$ ranges from 0 to 1 and a higher value corresponds to lower clustering coefficient. We set two groups of networks with $m$ equal to 3 and 9, indicating spare and dense networks respectively. For each group, we give three different values of temperature parameter $T$, as 0.1, 0.5 and 0.9, to generate networks with different average clustering coefficient. For each combination of the two parameters we generate 10 networks, and link prediction algorithms run 10 times for each realization. The average statistical features of their giant connected components are given in Table~\ref{t4}. 

\begin{figure*}
\includegraphics{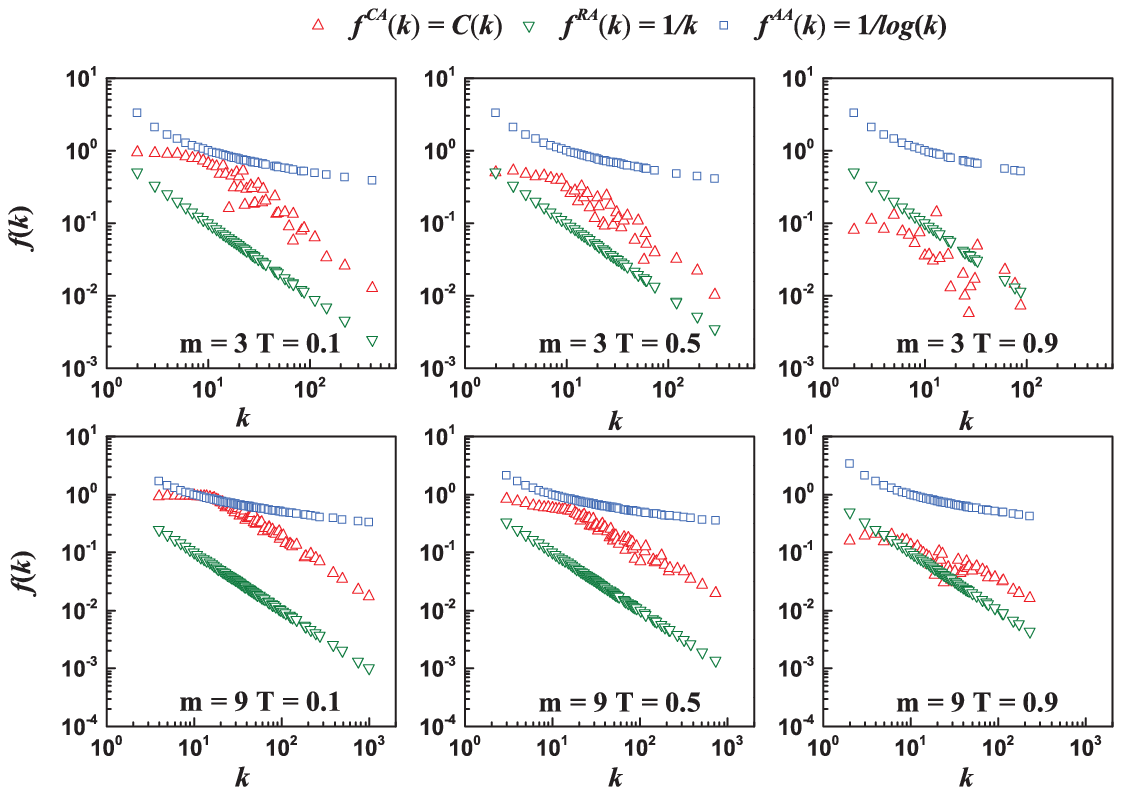}% Here is how to import EPS art
\caption{\label{fig2} Common-neighbor's contribution versus degree for CA, RA and AA in artificial networks. C(k) indicates the average clustering coefficient of nodes with degre k.}
\end{figure*}

We also plot $C(k)$, $1/k$ and $1/log(k)$ versus degree $k$ for the artificial networks in Figure~\ref{fig2}. Among these networks, the one with $m$ equal to 3 and $T$ equal to 0.9 is our most interested network, which is sparse and with low average clustering coefficient. More importantly, we find that the distribution of $C(k)$ versus degree is very similar to what we see in Figure~\ref{fig1}, i.e. for nodes with small degree, the trend of $C(k)$ is a bit ascending and for nodes with large degree, $C(k)$ distributes more broadly with the growth of degree $k$.

Table~\ref{t5} and Table~\ref{t6} show the link prediction results of four similarity indices on these artificial networks with different features. For results estimated under AUC, RA and CA almost give the same results, which are a little better than those of AA and CN. However when evaluated by precision, the big differences appear as what we see in real-world networks. Especially for the network with $m$ equal to 3 and $T$ equal to 0.9, CA index outperforms AA and RA by a large rate of $\bf{40\%}$ and $\bf{61.5\%}$, respectively. While on dense networks, differences among the four indices are very small. Thus what we find on the real-world networks are well verified by the test results on artificial networks generated by PS model. 

\begin{table}[h!]
\caption{The basic topological features of the giant component of artificial networks generated using PS model with different m and T. Other parameters of PS model are: N = 1000 (node number), $\zeta$ = 1 (curvature parameter), $\gamma$ = 2.1 (power law exponent).}
\begin{ruledtabular}
\begin{tabular}{llllll}
\multicolumn{1}{l}{\textbf{parameters}}&\multicolumn{1}{l}{\textbf{N}} & \multicolumn{1}{l}{\textbf{M}} 
&\multicolumn{1}{l}{\textbf{\textless k\textgreater}} & \multicolumn{1}{l}{\textbf{\textless d\textgreater}} & \multicolumn{1}{l}{\textbf{\textless C\textgreater}} \\ \hline
m=3 T=0.1&946.1&3160.8&6.684 &3.009 &0.783 \\  
m=3 T=0.5&969.7&3040.2&6.271 &3.212 &0.416 \\  
m=3 T=0.9&848.3&1459.9&3.442 &4.554 &0.071 \\  
m=9 T=0.1&1000&9720&19.440 &1.982 &0.851 \\  
m=9 T=0.5&1000&8912.9&17.826 &2.196 &0.532 \\  
m=9 T=0.9&993.9&4026.9&8.103 &3.029 &0.160 \\                                               
\end{tabular}
\end{ruledtabular}
\label{t4}
\end{table}

\begin{table}[h!]
\caption{Link prediction accuracy measured by AUC on artificial networks generated by PS model.}
\begin{ruledtabular}
\begin{tabular}{lllll}
\multicolumn{1}{l}{\textbf{AUC}}&\multicolumn{1}{l}{\textbf{CN}} & \multicolumn{1}{l}{\textbf{AA}} 
&\multicolumn{1}{l}{\textbf{RA}} & \multicolumn{1}{l}{\textbf{CA}}                                                \\ \hline

m=3 T=0.1&0.961&0.978&\bf{0.98}&\bf{0.98}\\  
m=3 T=0.5&0.859&0.872&\bf{0.873}&\bf{0.873}\\  
m=3 T=0.9&0.606&\bf{0.608}&\bf{0.608}&0.607\\  
m=9 T=0.1&0.986&0.994&\bf{0.996}&\bf{0.996}\\  
m=9 T=0.5&0.924&0.943&\bf{0.948}&\bf{0.948}\\  
m=9 T=0.9&0.712&\bf{0.72}&\bf{0.72}&\bf{0.72}\\

\end{tabular}
\end{ruledtabular}
\label{t5}
\end{table}

\begin{table}[h!]
\caption{Link prediction accuracy measured by precision on artificial networks generated by PS model.}
\begin{ruledtabular}
\begin{tabular}{lllll}
\multicolumn{1}{l}{\textbf{Prec}}&\multicolumn{1}{l}{\textbf{CN}} & \multicolumn{1}{l}{\textbf{AA}} 
&\multicolumn{1}{l}{\textbf{RA}} & \multicolumn{1}{l}{\textbf{CA}}                                                \\ \hline
m=3 T=0.1&0.482&0.663&\bf{0.687}&0.657\\  
m=3 T=0.5&0.174&0.2&0.193&\bf{0.211}\\  
m=3 T=0.9&0.036&0.03&0.028&\bf{0.041}\\  
m=9 T=0.1&0.962&0.984&\bf{0.986}&0.985\\  
m=9 T=0.5&0.352&0.363&0.369&\bf{0.37}\\  
m=9 T=0.9&\bf{0.101}&\bf{0.101}&0.091&0.099\\  

\end{tabular}
\end{ruledtabular}
\label{t6}
\end{table}

\subsection{Runtime}

At last, we show the efficiency of CA index in Table~\ref{t7}. Since the rest predicting procedures are the same for different similarity indices in the link prediction framework we used if similarity matrix is prepared, we only show the time cost of calculating the similarity matrix for the four indices. Clearly, CN runs fastest among the four indices. In most cases, CA can run competitively fast comparing with AA and RA. The most complex part of CA is the calculation of clustering coefficient, which has a computational complexity of $O(Nd_{max}^{2})$, where $d_{max}$ is the max degree of a network. Therefore, CA is very efficient, especially for sparse networks.

\begin{table}[h!]
\caption{Computing time (in millisecond) of similarity matrix for four similarity indices on 12 real-world networks. The hardware environment is the same for all similarity indices on the same network.}
\begin{ruledtabular}
\begin{tabular}{lllll}
\multicolumn{1}{l}{\textbf{networks}}&\multicolumn{1}{l}{\textbf{CA}} & \multicolumn{1}{l}{\textbf{CN}} 
&\multicolumn{1}{l}{\textbf{AA}} & \multicolumn{1}{l}{\textbf{RA}}                                                \\ \hline
PPI1&890.14 &69.68 &1223.93 &906.19 \\  
PPI2&1083.52 &81.74 &1398.48 &1060.61 \\  
Food&5.54 &0.14 &0.45 &0.40 \\  
Grassland&5.99 &0.11 &0.57 &0.48 \\  
Dolphins&6.51 &0.13 &0.49 &0.43 \\  
Jazz&32.14 &2.24 &4.62 &4.47 \\  
MacNeu&15.33 &1.03 &1.70 &1.65 \\  
MouseNeu&2.08 &0.08 &0.25 &0.24 \\  
PB&405.91 &44.25 &89.83 &70.74 \\  
Email&182.46 &10.13 &84.57 &79.04 \\  
Grid&1410.01 &64.26 &1185.65 &377.06 \\  
INT&833.25 &65.36 &2039.75 &393.49 \\  
\end{tabular}
\end{ruledtabular}
\label{t7}
\end{table}

\section{CONCLUSION}

In this paper, we present a novel efficient and parameter free common-neighbor based similarity index, called CA. The main difference from other well-known common-neighbor based similarity indices lies in the way of evaluating common-neighbor's contribution. CA index assumes common-neighbor with higher clustering ability contributes more to the likelihood of forming a link between a pair of nodes, and here average clustering coefficient of nodes with the same degree is used to measure the clustering ability of nodes with this degree. In both real-world networks and artificial networks, we find our measure of common-neighbor's contribution is very different with those of AA and RA, and the bigger the differences are, the better CA performs than AA and RA. 

Experimental results on both real-world networks and artificial networks show that CA index outperforms state-of-the-art common-neighbor based similarity indices in precision, especially on sparse networks with low average clustering coefficient. Although the calculation of clustering coefficient in CA index needs more time than AA and RA, the time costs of CA on real-world networks show that it is still a very efficient index. The computational complexity of clustering coefficient is $O(Nd_{max}^{2})$, where $d_{max}$ is the max node degree of a network. Thus, the computational complexity of CA is nearly $O(N)$ on sparse networks.

\begin{acknowledgments}
The authors acknowledges Linyuan L\"{u}, Tao Zhou and Carlo Vittorio Cannistraci for supporting some source codes and datasets used in this paper. This work is supported by the National Natural Science Foundation of China (Grants No. 61403023)
\end{acknowledgments}

\bibliography{clusteringAbility}% Produces the bibliography via BibTeX.

\end{document}